\documentclass[12pt,a4paper]{article}
\usepackage{epsfig}

\newcommand{\simge}
{\raisebox{-0.75ex}[-1.5ex]{$\;\stackrel{>}{\sim}\;$}}

\begin{document}

\title{Deformation and tribology of multi-walled hollow nanoparticles}
\author{U. S. Schwarz, S. Komura and S. A. Safran\\
Department of Materials and Interfaces,\\
Weizmann Institute of Science,\\
Rehovot 76100, Israel}
\maketitle

\begin{abstract}
  Multi-walled hollow nanoparticles made from tungsten disulphide
  (WS$_2$) show exceptional tribological performance as additives to
  liquid lubricants due to effective transfer of low shear strength
  material onto the sliding surfaces. Using a scaling approach based
  on continuum elasticity theory for shells and pairwise summation of
  van der Waals interactions, we show that van der Waals interactions
  cause strong adhesion to the substrate which favors release of
  delaminated layers onto the surfaces. For large and thin
  nanoparticles, van der Waals adhesion can cause considerable
  deformation and subsequent delamination. For the thick WS$_2$
  nanoparticles, deformation due to van der Waals interactions remains
  small and the main mechanism for delamination is pressure which in
  fact leads to collapse beyond a critical value. We also discuss the
  effect of shear flow on deformation and rolling on the substrate.
\end{abstract}

\section{Introduction}
Graphite and layered material made from metal disulphides (MoS$_2$,
WS$_2$) and similar composites (MoSe$_2$, WSe$_2$, BN, etc.) are good
lubricants since under shear the layers can easily slide over each
other due to atomic smoothness and weak van der Waals (vdW)
interactions \cite{b:sing92}.  Therefore they are widely used as solid
lubricants or additives to liquid lubricants.  However, finite sized
crystallites in powders have edges with dangling bonds which can react
chemically with the sliding surfaces.  This problem can be avoided by
using hollow nanoparticles made from the same material, where the
layers are not planar, but are bent into closed shells. It has been shown
recently that multi-walled WS$_2$ nanoparticles perform very well as
additives \cite{n:rapo97,n:rapo99}. Closer investigations using the
Surface Force Apparatus (SFA) revealed that WS$_2$ layers are
transfered onto the sliding surfaces by delamination of the
nanoparticles \cite{n:gola99}. These peeled layers form islands on
which the Friction Force Microscope measured much smaller friction
than on the surrounding mica. Hence multi-walled nanoparticles can act
as reservoirs which release layers of low shear strength exactly where
needed while avoiding too many dangling bonds.

\begin{figure}[h]
  \begin{center}
    \epsfig{file=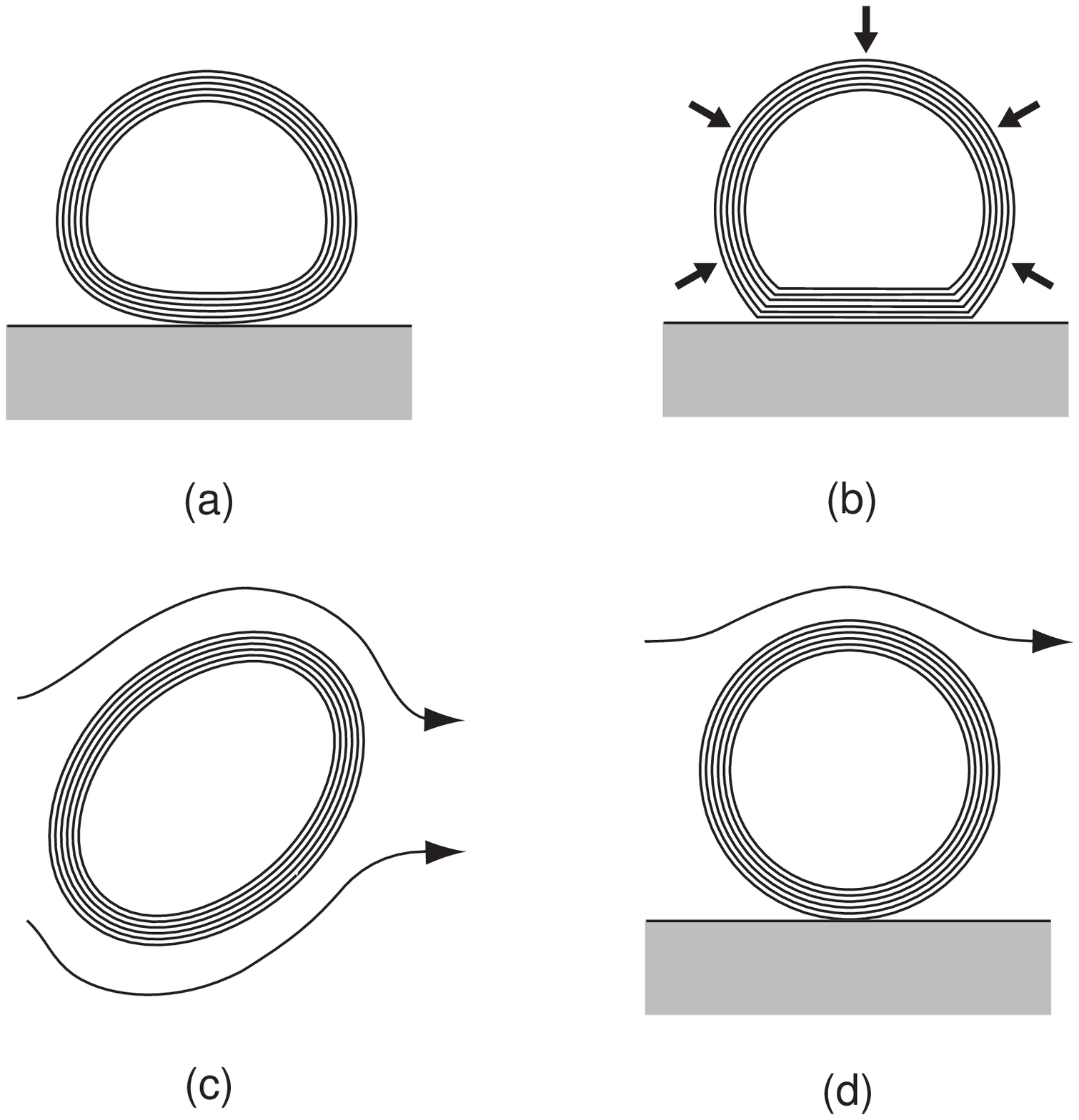,width=\textwidth}
  \end{center}
  \caption{Several aspects of friction experiments with multi-walled nanoparticles:
    (a) weak deformation due to adhesion onto a substrate, (b) strong
    deformation due to pressure, (c) deformation in shear flow and (d)
    rolling of adhering particles in shear flow.}
  \label{overview}
\end{figure}

A theoretical understanding of the mechanical properties of
multi-walled hollow nanoparticles is important for future tribological
applications. In this paper, we use scaling arguments based on
continuum elasticity theory and pairwise summation of vdW interactions
in order to investigate theoretically the effect of adhesion, pressure
and shear flow on the deformation and mechanical stability of hollow
nanoparticles of spherical shape. Figure \ref{overview} schematically
depicts some of the aspects discussed in the following.  We show that
vdW adhesion to the substrate (as well as to each other) can be
several orders of magnitude larger than thermal energies and scales
linearly with radius, but is almost independent of thickness.  Despite
the large energy of adhesion, we find that for typical WS$_2$
nanoparticles, coherent and even incoherent deformations due to the
vdW adhesion to the substrate are only in the Angstrom range and will
leave the nanoparticles basically intact. Thus for WS$_2$
nanoparticles the vdW adhesion favors the release of delaminated
layers onto the surfaces, but does not trigger the delamination
itself.  The main mechanism for delamination of WS$_2$ nanoparticles
is shown to be pressure which leads to a mechanical instability in
linear elasticity theory. This might explain why damage of WS$_2$
nanoparticles was found to occur only beyond a critical load
\cite{n:rapo99}. We also show that shear flow does not lead to
considerable deformation of WS$_2$ nanoparticles and that very large
shear rates are needed in order to make adhering particles roll in
shear flow.

Similar methods have been used before to predict the shape of
fullerenes \cite{n:ters92,n:witt93} and to account for the faceted
shape of metal disulphide particles \cite{n:srol94,n:srol95}.
Compared with {\it ab initio} methods \cite{n:adam92}, tight-binding
schemes \cite{n:hern98} and molecular simulations \cite{n:yako96}
which have been used before to investigate the mechanical properties
of fullerene-like material, our approach has the advantage that it is
asymptotically correct for large systems and universal in the sense
that different material systems enter on the level of their elastic
and vdW constants. In particular we discuss C, MoS$_2$ and WS$_2$,
although we focus on the WS$_2$ nanoparticles used in the tribological
experiments mentioned above.  Our scaling approach allows us to
predict how tribological performance depends on radius $R$, shell
thickness $h$ and layer thickness $a$ of the nanoparticles.  We show
that scaling with $h$ does not result from the vdW contributions, but
rather from the scaling of the elastic constants with $h$. Several
critical quantities derived scale strongly with $R/h$ for coherent
deformation and with $R/a$ for incoherent deformation. Hence
tribological performance can be tuned by varying the concentration of
defects (which switches between the coherent and incoherent regime)
and the ratios $R/h$ and $R/a$, respectively.

\section{Preliminaries}
In the following we consider hollow nanoparticles with outer radius
$R$ and thickness $h$ which results from nesting several elastic
shells of thickness $a$ each. Typical WS$_2$ particles used in
tribological experiments have $R \approx 60$ nm and $h \approx 9.3$ nm
($15$ layers with an interlayer distance of $a = 0.62$ nm, $R / h
\approx 6$, $R / a \approx 100$). In fig.~\ref{tem} a high resolution
transmission electron micrograph of a WS$_2$ nanoparticle with $11$
layers is shown.  The elastic shells are closed since their formation
is driven by the energy reduction due to the absence of dangling
bonds. In order to achieve closure, topology dictates that for carbon
shells $12$ pentagons have to be inserted into the graphite network of
hexagons. For hexagonally layered composites, the sheets have a more
complicated molecular structure (typically triple layers) and
different kinds of defects have to occur to ensure closure. In our
continuum approach, we assume that defects are distributed in a
homogeneous way to give a spherical shape with a certain preferred
radius $R$ \cite{n:srol94}. Similar approaches to fullerene-like
particles have considered sheets which are planar in equilibrium; in
order to account for their curvature, they explicitly considered
defects \cite{n:ters92,n:witt93,n:srol94,n:srol95}.  However, in this
approach the elastic response of a spherical shell is very complicated
and no unifying scaling approach is possible.

The new feature of the elasticity of a closed shell with a preferred
radius $R$ is that stretching is a first-order effect and the
spherical shell cannot be bent without being stretched
\cite{b:land70}.  This interplay between bending and stretching has
been studied before for thermal fluctuations of polymerized vesicles
\cite{e:komu92}.  The elastic behavior of the shell is determined by
two contributions: bending energy $E_b \sim \kappa \int dA\ c^2$ with
bending rigidity $\kappa$ and mean curvature $c$, and stretching
energy $E_s \sim G \int dA\ e^2$ with in-plane stretching modulus (or
two-dimensional Young modulus) $G$ and in-plane strain $e$. Here $\int
dA$ represents the surface integral; the elastic contributions in
shell theory follow by integrating over the thickness of the shell.
Consider a spherical shell of radius $R$ which is expanded by $\Delta
R$. Then $c$ changes by $\Delta R / R^2$ and $e$ by $\Delta R / R$,
thus both contributions are first order effects.  Since $E_b / E_s
\sim \kappa / G R^2$, the relative strength of bending and stretching
depends both on material parameters and radius; on large length scales
$R \gg (\kappa / G)^{1/2}$ and stretching will always dominate.

In the framework of continuum elasticity theory, the elastic moduli
$\kappa$ and $G$ can be calculated from the in-plane stretching
elastic constant $C_{11}$ of the corresponding hexagonal layered
material as $\kappa = C_{11} h^3 / 12$ and $G = C_{11} h$
\cite{n:srol94}. The values for $C_{11}$ are $1060$, $238$ and $150
\times 10^{10}$ erg/cm$^3$ for C, MoS$_2$ and WS$_2$, respectively
\cite{b:land91}.  The same scaling with $h$ is found for thin films
made from isotropic elastic material \cite{b:land70}. It only applies
for coherent bending of the different layers; for incoherent bending,
slip occurs between adjacent layers and overall bending becomes
easier. The bending rigidity then follows as $\kappa \sim C_{11} a^3
(h / a) = C_{11} a^2 h$ where $a$ is the effective thickness of a
single layer \cite{n:srol94}.  It should be of the same order of
magnitude, but somewhat smaller than the interlayer distance (which is
$3.4\ \AA$ for C and $6.2\ \AA$ for MoS$_2$ and WS$_2$). Note that the
in-plane modulus $G \sim C_{11} a (h / a) = C_{11} h$ stays the same
since it scales linearly with thickness.  The change from coherent to
incoherent bending can be considered to be defect mediated;
dislocations will proliferate for increasing thickness and finally
will lead to grain boundaries which account for the typical faceted
shape seen in fig.~\ref{tem} \cite{n:srol94}.

\section{Adhesion}
We consider a substrate which interacts by attractive vdW forces with
a film of thickness $h$ which is a distance $D$ away from the
substrate.  Pairwise integration of the potential $-A / \pi^2 r^6$
(where the Hamaker constant $A$ is typically of the order of $10^{-12}$
erg) over the two volumes yields the vdW energy per unit area
\begin{equation}
  \label{eq:vdW}
  u = \frac{A}{12 \pi} \frac{h (h + 2 D)}{D^2 (h + D)^2}\ .
\end{equation}
Here $D$ is an atomic cutoff for the vdW interaction which in the
following is chosen to be $1.65\ \AA$ \cite{a:isra92}.  Although the
adhesion energy $u$ scales linearly with $h$ for small $h$, for more
than one layer we have $h > D$ and $u$ saturates at a constant value
$u = A / 12 \pi D^2 \approx 100$ erg/cm$^2$.  For the case of a
single layer, one can replace the volume integral over the film by a
surface integral times the effective thickness $h$; this is equivalent
to considering $h \ll D$ in eq.~(\ref{eq:vdW}) and leads to $u = A h
/ 6 \pi D^3$. Since $D$ and $h$ are of the same order of magnitude, we
essentially recover the result for a multi-layered film.  Hence we
conclude that due to the rapid saturation of the vdW binding energy
with increasing $h$, for multi-layered films under adhesion conditions
no considerable scaling with $h$ is expected from the vdW terms.

It can be shown by similar arguments ({\it e.g.}\ by using
eq.~(\ref{eq:vdW}) within the Derjaguin approximation) that the energy
of a hollow nanoparticle adhering to a substrate does not depend on
$h$ as long as $h \simge D$. Then the adhesion energy equals the vdW
energy of close approach between a sphere and a substrate, $E_A = A R
/ 6 D$ \cite{a:isra92}.  Using values for $A$, $D$ and $R$ as given
above yields $E_A = 6 \times 10^{-11}\ \mbox{erg} = 1400$ kT.  The
energy of adhesion between two particles is only a factor of $2$
smaller than the one between a particle and the substrate, and hence
well above thermal energies as well. We thus conclude that the vdW
interaction leads to considerable adhesion of the particles to the
substrate and to each other. The same holds true for peeled off outer
layers: with $R = 60$ nm, the energy of adhesion can be estimated to
be $R^2 u = 10^5$ kT.

\section{Deformation under adhesion}
\begin{figure}[h]
  \begin{center}
    \epsfig{file=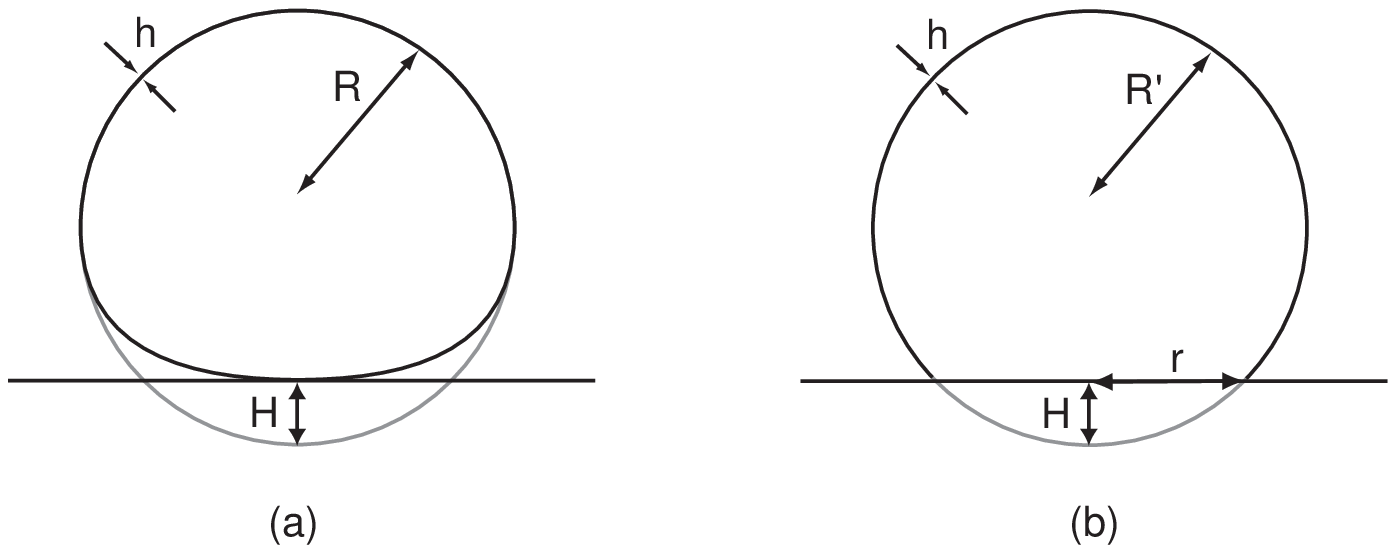,width=\textwidth}
  \end{center}
  \caption{Possible deformations of a spherical elastic 
    shell on contact with a substrate: (a) flattening and (b) contact
    disc.  $R$ is the initial particle radius, $R'$ the radius after
    flattening, $h$ the shell thickness, $H$ the indentation and $r$ the
    radius of the contact disc. For small $H$, one finds geometrically
    $r^2 = 2 H R'$.}
  \label{deformation}
\end{figure}

The onset of delamination can be estimated by considering large
deformations of the nanoparticles. We do not offer a theory for
fracture and estimate the limit of mechanical stability by an internal
criterion within our continuum description: fracture and delamination
sets in when so much stress has accumulated that the deformation
becomes of the order of the radius.  Determining the deformation of a
spherical elastic shell is a difficult problem which depends on
prestress, elastic constants and size as well as on the nature of the
deforming force. Here we discuss the deformation on a substrate for
two cases which can be treated in the framework of our scaling
approach. In the following $H$ is the indentation. For small
deformations $H < h$ (fig.~\ref{deformation}a), the shell flattens at
the bottom. For large deformations $H > h$ (fig.~\ref{deformation}b),
a contact disc with radius $r$ develops and the elastic energy is
localized in the circular fold surrounding it. The respective elastic
energies are \cite{b:land70}
\begin{equation}
  \label{eq:deformation}
  E_{small} \sim \frac{G^{1/2} \kappa^{1/2}}{R} H^2,\ \quad
  E_{large} \sim \frac{G^{1/4} \kappa^{3/4}}{R} H^{3/2}\ .
\end{equation}
Here we neglect a bending term $\sim \kappa H / R$.  We also assumed
that due to low friction, slip can occur between the contact disc and
the substrate; otherwise compression energy $\sim G H^3 / R$ would
accumulate in the contact disc. The rest of the relaxing shell is
assumed to keep a spherical form with radius $R' > R$ in order to keep
area constant and to avoid compression energy.  However, the
difference is $(R' - R) \sim H^2 / R$ and can be shown to be
negligible in the following.

If the deformations are driven by vdW adhesion, the energy gained on
adhesion can be shown to scale as $E_A \sim u R H$ for both small and
large deformations. Setting this adhesion energy equal to the estimates
for the elastic energies from eq.~(\ref{eq:deformation}) yields
estimates for indentation:
\begin{equation}
  \label{eq:flattening}
  H_{small} \sim \frac{R^2 u}{G^{1/2} \kappa^{1/2}} 
\sim \left( \frac{u}{C_{11} h} \right) \left( \frac{R}{h} \right) R,\ \quad
  H_{large} \sim \frac{R^4 u^2}{G^{1/2} \kappa^{3/2}} 
\sim \left( \frac{u}{C_{11} h} \right)^2  \left( \frac{R}{h} \right)^3 R\ .
\end{equation}
Here and in the following we will always give the result both in terms
of the elastic moduli and their scaling with thickness for coherent
bending.  The dimensionless quantity $u / C_{11} h$ is the ratio of
van der Waals adhesion energy to two-dimensional Young modulus and
will be of order $10^{-4}$ for the WS$_2$ nanoparticles.  In this case
we are in the regime $H < h$ and deformations will be in the Angstrom
range. Thus delamination cannot be expected to be caused by adhesion.
For incoherent bending, $R/h$ has to be replaced by $R/a$ in both
regimes.

One system for which vdW adhesion leads to strong deformations are
{\it single-walled} hollow nanoparticles.  In this case, care has to
be taken to use the correct values for $G$ and $\kappa$.  Since $G$ is
a purely two-dimensional quantity, $G = C_{11} h$ can be used where $h$
is identified with the interlayer distance. However, $\kappa$ has to
be extracted from molecular calculations
\cite{n:ters92,n:hern98,n:yako96}. For carbon, one finds $G = 3.6
\times 10^5$ erg/cm$^2$ and $\kappa \approx 1.6 \times 10^{-12}\ 
\mbox{erg} \approx 40$ kT. For both MoS$_2$ and WS$_2$, $G$ is smaller
by a factor $4$ and $\kappa$ is larger by a factor $10$.  In all three
cases, for radii larger than a few nm considerable deformations arise.
The critical radius $R_c$ where indentation $H$ and radius $R$ become
of the same magnitude follows from the second relation of 
eq.~(\ref{eq:flattening}) as $R_c \sim
G^{1/6} \kappa^{1/2} u^{-2/3}$.  For C, we find $R_c \approx 5\ nm$
and for the metal disulphides $R_c \approx 10\ nm$.  
For nanotubes, we find similar results which are in good agreement 
with experiments and molecular calculations \cite{n:ruof93,note}.

\section{Deformation under pressure}
We now proceed to show that pressure can lead to an instability which
could account for the observed delamination.  We consider the
formation of a contact disc for a shell which is pressed onto the
substrate by pressure in the surrounding liquid. The corresponding
energy is $E_p \sim -p R H^2$, thus it scales more strongly with $H$
than the restoring elastic energy $E_{large}$ in
eq.~(\ref{eq:deformation}). This scaling indicates an instability:
small deformations are suppressed, but large deformations grow without
limits beyond the critical indentation $H_c$ where $E_{large}+ E_p$
attains a maximum \cite{b:land70}.  Setting $H_c = R$ gives an
estimate for the critical pressure for delamination:
\begin{equation}
  \label{eq:criticalpressure}
  p_c \sim \frac{G^{1/4} \kappa^{3/4}}{R^{5/2}} 
\sim C_{11} \frac{h}{R} \left( \frac{h}{R} \right)^{3/2}\ .
\end{equation}
Using typical values gives $p_c \approx 1.7$ GPa. Incoherent
deformations can be treated by replacing $(h/R)^{3/2}$ by
$(a/R)^{3/2}$ in eq.~(\ref{eq:criticalpressure}) and hence would
decrease the estimate to $25$ MPa.  Although a full treatment of the
instability requires a theory for the non-linear elasticity, we can
still suggest that pressure will be a very likely mechanism to cause
delamination. This prediction agrees nicely with the experimental
observation that damage sets in at a finite threshold of the load
\cite{n:rapo99}.  Note that for single-walled buckyballs C$_{60}$ with
$R = 3.55\ \AA$, eq.~(\ref{eq:criticalpressure}) gives $p_c = 15$ GPa
which is in surprisingly good agreement with the value $20$ GPa found
experimentally \cite{n:nune92}.

Although the pressure values estimated here seem to be high, they
refer to particles of nm-size, and the corresponding forces are of the
order of $10^{-3}$ dyn.  In friction experiments, the pressures and
forces needed for delamination are most likely to occur near
asperities.  This might explain why delamination has been observed in
the SFA only when shearing the two surfaces
\cite{n:gola99}: the nanoparticles then behaves as granular material
which gets jammed and pressure and forces become localized.

\section{Shear flow}
Since the WS$_2$ nanoparticles are immersed in liquid lubricant in
strong motion, we finally consider the effect of shear flow.
Balancing the deforming viscous force with the elastic restoring force
and setting the resulting deformation equal to the radius gives an
estimate for the shear rate at which delamination sets in. For $h <
R$, we find that only stretching is relevant and that the critical
shear rate is
\begin{equation}
\label{shearflow}
\dot\gamma_c \sim \frac{G}{\eta R} \sim \frac{C_{11}}{\eta} \frac{h}{R}\ .
\end{equation}
Using viscosity $\eta \approx 1$ cP (which applies both to water and
mineral oil), we find $\dot\gamma_c \sim 2.5 \times 10 ^{13}$ Hz.  
Such high shear rates are unattainable even near contacting asperities in
macroscopic friction experiments, thus delamination is not likely to
occur due to shear flow alone.

Shear flow past nanoparticle adhering to a substrate can cause them to
roll. The particle begins to roll if the viscous drag evaluated at the
midpoint, $F_S = 6 \pi \eta R^2 \dot\gamma$, equals the friction force
$F_F$ caused by the adhesive load $F_A = 2 \pi R u$. Assuming
Amontons' law for rolling friction, $F_F = \mu F_A$, we find
\begin{equation}
  \label{eq:rolling}
  \dot\gamma_c = \frac{\mu u}{3 \eta R}
\end{equation}
where the coefficient of rolling friction $\mu$ can be taken to have
the typical value $10^{-3}$; then $\dot\gamma_c$ is of the order of
$5 \times 10^5$ Hz. This value is higher than typical shear rates probed
with the SFA ($v = 10\ \mu$m/s, $\dot\gamma = 10^4$
Hz on nm-separation), but certainly lower than shear rates occurring
near the asperities of macroscopic friction experiments ($v = 1$m/s, 
$\dot\gamma = 10^9$ Hz on nm-separation). 

\section{Conclusion}
We have shown that in friction experiments with multi-walled hollow
nanoparticles, delamination is likely to be caused by pressure. When
nanoparticles delaminate, dangling bonds develop as they do for
powders from layered material. However, there are two reasons why they
should be less problematic. First delamination takes place with one or
two layers and the density of dangling bonds remains low. Second we
showed that delamination may occur under adhering conditions. The
adhesion (and possibly the shear) will cause the layers to align
parallel to the substrate, in contrast to the powders where the
orientation is preferentially perpendicular to the substrate. It has
been shown experimentally that delaminated layers from nanoparticles
form islands while platelets form a thicker, more disorganized film
\cite{n:gola99}. Our scaling approach shows that there are two control
parameters which might be used for tribological optimization: the
ratio of radius to thickness and the concentration of defects (that
determine the regimes of coherent and incoherent bending).

Using nanoparticles as additives to liquid lubricants raises
interesting questions about the role of wetting. The lubricant used in
the friction experiments mentioned is a bad solvent for the
nanoparticles, which provides an additional force to drive them onto
the substrate. A dewetted region is expected to form between the
adhering particle and the substrate; like a capillary bridge in a good
wetting situation, it might enhance adhesion.

\section*{Acknowledgments}
\begin{figure}[h]
  \begin{center}
   \epsfig{file=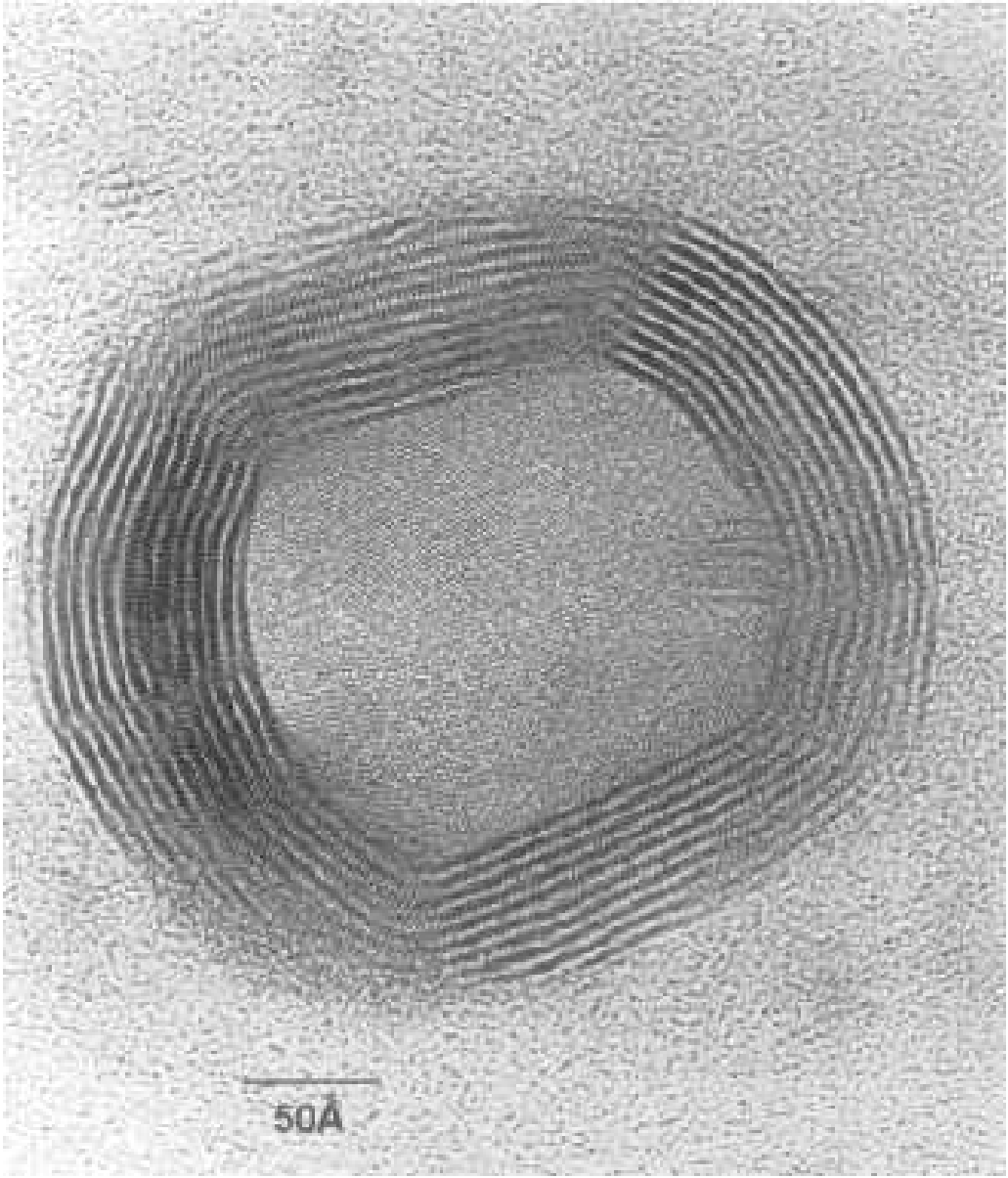,width=\textwidth}
  \end{center}
  \caption{High resolution transmission electron micrograph 
    of a multi-walled WS$_2$ nanoparticle.  The slightly faceted shape is
    typical and can be attributed to the assembly of defects into
    grain boundaries \protect\cite{n:srol94}.}
  \label{tem}
\end{figure}

We thank {\sc C. Drummond}, {\sc Y. Golan}, {\sc J.
  Israelachvili}, {\sc J.  Klein}, {\sc G. Seifert} and
especially {\sc R. Tenne} for helpful discussions. {\sc R.
  Tenne} kindly provided the electron micrograph shown in
fig.~\ref{tem}.  {\sc SAS} thanks the Gerhard M.J. Schmidt Minerva
Center for Supramolecular Architecture for support. {\sc SK} 
thanks the Ministry of Education, Science and Culture, Japan
for providing financial support during his visit to Israel.
{\sc USS} gratefully acknowledges support by the Minerva
Foundation.

\begin{thebibliography}{10}
\bibitem{note} 
Our calculation for nanotubes is not given here since deformations of
cylindrical shells are conceptually different from the ones of
spherical shells: for spherical shells, the balance of stretching
and bending leads to localization of deformation. For cylindrical
shells, stretching can be avoided and non-local bending has to be
considered.
\end{thebibliography}
\end{document}